# Detecting Forged Kerberos Tickets in an Active Directory Environment

Thomas Grippo, Hisham A. Kholidy
State University of New York (SUNY) Polytechnic Institute, Network and Computer Security Department, Utica, NY USA
grippot@sunypoly.edu, hisham.kholidy@sunypoly.edu

**Abstract-** Active Directory is the most popular service to manage users and devices on the network. Its widespread deployment in the corporate world has made it a popular target for threat actors. While there are many attacks that target Active Directory and its authentication protocol Kerberos, ticket forgery attacks are among the most dangerous. By exploiting weaknesses in Kerberos, attackers can craft their own tickets that allow them to gain unauthorized access to services on the network. These types of attacks are both dangerous and hard to detect. They may require a powerful centralized log collecting system to analyze Windows security logs across multiple services. This would give additional visibility to be able to find these forged tickets in the network.

## I. INTRODUCTION

As the world becomes more reliant on technology, the impact of cyberattacks will continue to grow. Many people began to realize this in March of 2020 when the COVID-19 pandemic began. This was an unprecedented time as the world faced widespread closures and whole economies came to a grinding halt. The dangers of this highly transmissible disease forced employers to send everyone home, and thus the work from home culture began to take off. This shift made people very dependent on their computers and a reliable Internet connection as they had to constantly attend meetings and Zoom calls from their couch. Relying on technology more than ever, companies and their employees could not afford to face the ramifications of a widespread cyber-breach and thus had to invest a lot of time and money into protecting the infrastructure that made these new workflows possible.

Two years later, another unprecedented event caused a greater focus on cybersecurity. In February of 2022, Russia began its invasion of Ukraine. Being how interconnected the major countries were in the world economy, nobody expected there to be another large-scale invasion such as this, and its effects were felt around the world. More interestingly, it was the first major conflict where cyberwarfare became a prominent factor. One DNS platform claimed that they blocked 10 times the normal number of malware attacks targeting people in Ukraine [1]. There were also major disruptions in Internet traffic as Russia cut power in many Ukrainian cities. Due to the nature of cyberwarfare, individuals around the world also were able to play a part in the conflict, going after Russian infrastructure in support of Ukraine.

Looking at the current state of the world, the importance of defending cyberspace is becoming clearer to everyone. Modern assets will continue to thrive on the Internet, and for them to function properly they must be defended. This is true for any asset whether it be government or corporate-owned.

One of the biggest issues many organizations face is managing all the assets they have on the network. To help solve this, Windows Active Directory is employed. Windows Active Directory, or AD for short, is a directory service that provides methods to store directory data and make it available to network users and administrators. This directory data can be any information about a certain object such as a user or service [2]. In essence, Active Directory acts as a centralized database for all the users and services that exist within a network, and it is invaluable for administrators managing this environment. In fact, it is stated that over 90% of the Global Fortune 1000 companies, that is the top 1000 companies, use AD domain services [3].

With such widespread dominance in the corporate world, Active Directory has become a target for many threat actors. Specifically, they have targeted AD's default authentication service Kerberos. Kerberos is what allows users within an AD environment to access resources, and thus if it can be exploited by an attacker, that attacker would also be able to access such resources. This can be very troubling because if an attacker can exploit the Kerberos protocol, they can easily pivot to any part of the network and maintain persistence. That is why securing Active Directory, and detecting attacks within the environment should be a priority for IT.

In this paper, we will be focusing on two different attacks on Kerberos, the Golden Ticket attack, and the Silver Ticket attack, which can be collectively referred to as forged Kerberos ticket attacks. We will explore how these attacks function within a virtualized Active Directory environment, and more importantly how to detect them. The rest of this paper will be structured as follows: Section II will provide the necessary background information to understand this experiment. This will include an in-depth look at Windows Active Directory, Kerberos Authentication Services, and the Golden Ticket and Silver Ticket attacks. Section III will give an overview of the virtual environment that will be set up to carry out these experimental tests. Then, Section IV will walk through the operation of the Kerberos attacks within the virtual environment and how they can be detected. Section V will go over the results of this experiment and discuss the potential impacts in the real world. Finally, Section VI will provide a conclusion to everything discussed in this paper.

## II. BACKGROUND

This section will provide an overview of the background knowledge needed to understand the components of this experiment. Topics such as Windows Active Directory and the Kerberos Authentication Protocol will be explained in detail.

### A. Active Directory Domain Services

As stated previously, almost all modern organizations are using Windows Active Directory to manage the users in their network, so it is important to understand this Microsoft product in depth.

At the very core of Active Directory lies the users and the computers. People and their computers are what fundamentally make up a network, and in AD they will each have their own account. Shown below in *Figure 1* is an example user account in AD.

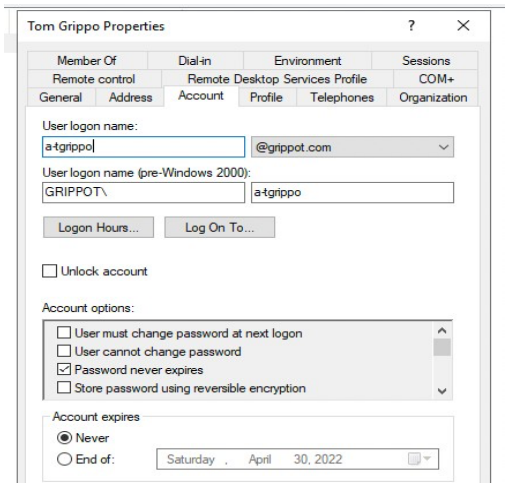

*Figure 1: User Account in Active Directory*

The user login name is shown, as well as the domain that it belongs to which is *grippot.com*. Instead of individually assigning permissions to each user account, they can instead be added to security groups within AD. These security groups can then be assigned specific permissions. For every user account, we can see what security groups they are a part of under the *Member Of* tab shown in *Figure 2*. As shown, this user is part of the *Domain Admins* and *Domain Users* security groups, which are default groups created when Active Directory is configured. Each of these groups has a set of permissions defined that will apply to any users that are a part of the group. So for instance, this user will have all of the permissions of a *Domain Admin*.

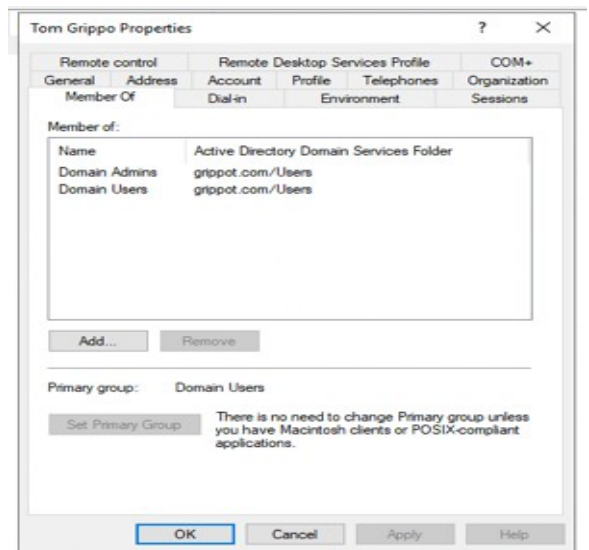

*Figure 2: Security Groups in Active Directory*

This streamlines the process of access control. For example, if a new employee gets hired within the accounting department, instead of manually assigning them all the permissions they need, you can simply add them to a custom accounting security group and they will inherit all of their permissions.

So far we have user and computer accounts, as well as security groups to manage permissions. If you want to organize all of these accounts within Active Directory, you can create organizational units (OUs) or containers as shown in *Figure 3*.

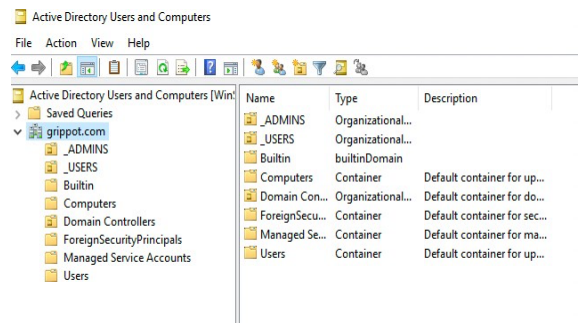

*Figure 3: OUs and Containers*

In this screenshot, the containers are plain folders and the OUs are marked. The main difference being that you can apply group policy to OUs but not to containers [2]. So if you want to organize all of the users, you can create OUs based on the different departments within the company.

Taking a step back, all these objects discussed exist within a domain. This can be seen in *Figure 3*, where all the OUs and containers, as well as the user and computer accounts that exist within them, belong to the *grippot.com* domain. The domain is controlled by the domain controller which is typically a Windows server. The domain controller is where all the objects are stored and can be managed.

For the purposes of this paper, this is all the information that is needed. It should be noted though, that Active Directory does not stop with domains.
There are also trees (a domain and its subdomains) as well as forests. However, this will not be relevant to the experiments in this paper.

### B. Kerberos

Another component of Active Directory is the Kerberos Authentication Protocol [4]. This is what allows a user account to access a computer or resource within a domain. It performs both authentication and authorization based on the permissions of the user account. In an AD environment, the domain controller will act as the Key Distribution Center or KDC. The KDC is made up of the Authentication Service (AS), and the Ticket Granting Service (TGS).

Shown below in *Figure 4* is an overview of the message exchange within Kerberos, we will use this diagram to analyze each step in the Kerberos protocol. In this example, we will say that a user is logging on to a desktop instead of accessing a server as shown in *Figure 4*.

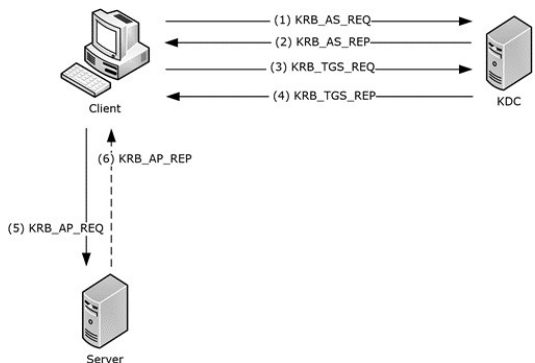

*Figure 4: Kerberos Message Exchanges* [5]

For a user to log on to a computer, they must enter their username and password. Once they hit enter, the Kerberos message exchange will begin. First, the client will send a KRB_AS_REQ to the AS on the KDC. This stands for Kerberos Authentication Service Request and is shown below in *Figure 5*.

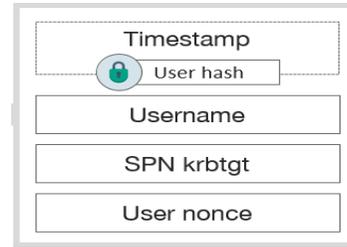

*Figure 5: KRB_AS_REQ* [6]

This message will include the username, a timestamp encrypted with the user's password, and the service principal name of the requested service. Since the user is requesting authentication, this will be the SPN of the krbtgt account, a default account within AD [5]. Once this message is sent to the KDC, it will look at the username inside the message. It will then retrieve the associated password key with that account and attempt to decrypt the timestamp. If it can decrypt the timestamp, then the user's identity has been proven and authenticated. It will also verify that the timestamp is not too old.

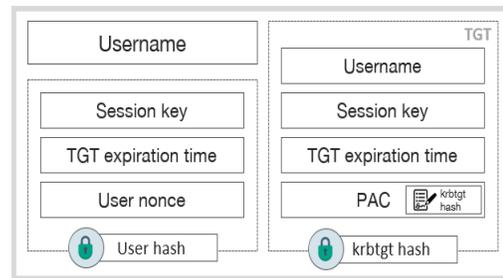

*Figure 6: KRB_AS_REP* [6]

The next message shown above in *Figure 6* is the response of the KDC. Since the user sent the SPN of the krbtgt account in the prior message, it knows that the user wants a ticket-granting ticket (TGT). This TGT acts as proof of authentication, and will be used to request tickets for specific services within AD. As shown above, the TGT is encrypted with the hash of the krbtgt account, and it includes the username of the user, a session key that will be used for future message exchanges, the TGT expiration time, and the PAC which includes all the user's permissions. The KRB_AS_REP message also includes the username and the same session key which is encrypted with the user's password key. When this message is received by the user, they will not be able to decrypt the TGT, but instead will make a copy and store it in memory for future use. They will be able to decrypt the rest of the message which includes a session key. This is a symmetric key that will be used to encrypt future messages to the KDC.

At this point, the user has successfully authenticated and has a TGT. Now they will use this TGT to get a service ticket for the computer they are trying to log in to.

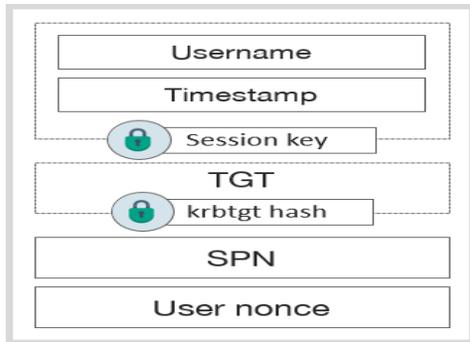

Figure 7: KRB_TGS_REQ [6]

First, they will send a ticket-granting service request or KRB_TGS_REQ to the KDC. Take note that the KDC is now acting as the TGS instead of the AS. The user will copy the TGT from RAM into the message as proof of their authentication. They will also place their username and timestamp, which will be encrypted with the session key that they received in the prior message. Also, the SPN of the computer they are attempting to log into will be placed inside the message unencrypted. When the KDC receives this message, it will be able to decrypt the TGT with the krbtgt hash. Inside the TGT, it will retrieve the session key to decrypt the rest of the message. It will verify the timestamp and expiration time of the TGT. Then, it will begin to craft the KRB_TGS_REP message.

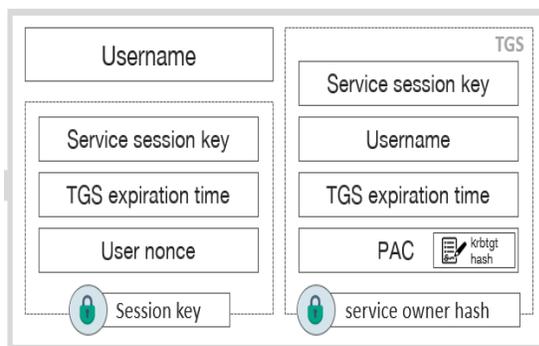

Figure 8: KRB_TGS_REP [6]

The KDC will find the hash of the requested service based on the SPN in the prior message. It will then generate another session key that will be used in future communications between the client and the service. It will take the key, and the username with its associated permissions, and will encrypt it with the hash of the service account. This will be the TGS ticket that the user uses to access the service. The KDC will also make another copy of the service session key for the user, which will be encrypted with the user's hash. When the user receives this message, they will copy the TGS ticket into RAM. They will also decrypt the rest of the message so that they can access the service session key.

So far the user has authenticated with the KDC in order to get a TGT, and they used their TGT to get a TGS for the computer they are trying to access. Now they can begin to communicate with the computer.

First, they will send a KRB_AP_REQ message to the service, as shown below in *Figure 9*.

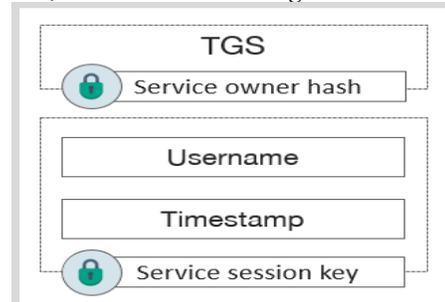

Figure 9: KRB_AP_REQ [6]

The TGS will be copied from RAM and placed inside the message along with the username and a timestamp which is encrypted with the service session key. This message will then be sent to the service which in this case is the same computer the user is trying to log in to. When the service receives this message, it will attempt to decrypt the TGS with its hash. If it is successful, it knows that this ticket came from the KDC since the KDC, or domain controller is the only other server that has its hash. Inside the TGS, the service will retrieve the service session key to decrypt the rest of the message. It will then check the username of the user and verify that they have valid permissions. Recall that in the KRB_TGS_REP message, the KDC included the user's permissions within the TGS. After this verification step, the service will decide whether to grant the user access. Assuming it will, the service will send a KRB_AP_REP back to the user and they will be able to login to the computer.

After the user is logged into the computer, they can then access other resources within the Active Directory Domain. Each time they want to access a different service, they will need to retrieve a TGS from the KDC and repeat the same process.

*C. Attacks on Kerberos*

Now that Active Directory and Kerberos have been explained, it will be easier to understand the various attacks against the Kerberos protocol. Since Kerberos is a secure protocol, many of these attacks are advanced. For the purposes of this paper, we will only be looking at two types of attacks, Golden Ticket, and Silver Ticket attacks.

*1) Silver Ticket Attack:* The Silver Ticket attack is when an attacker forges a TGS ticket. For this to be successful, the attacker would need the password or NTLM service hash of the service account to be able to properly encrypt the TGS ticket [7]. They could then make up any session key they want to use, and add in the rest of the information inside the TGS ticket such as the expiration time and the PAC. For the PAC, ideally, the attacker would give themselves admin privileges so they can have full access to the service.

*2) Golden Ticket Attack:* The Golden Ticket attack is when an attacker forges the TGT. For this to be successful, the attacker would need the hash of the krbtgt account, which is much harder to acquire than any service account hash [7]. However, if an attacker is successful, they can essentially impersonate the KDC and give themselves administrator rights to any service in the domain. This attack is extremely dangerous if successful, but it is harder to perform and noisier than the silver ticket attack.

### III. THE VIRTUAL ENVIRONMENT

To see these Kerberos attacks in real-time and attempt to detect them, we will need to set up a virtual environment. Shown in *Figure 10* is a diagram of the environment that has been set up for this experiment. To keep things simple, there are only three systems in the *grippot.com* domain, the domain controller, a Windows 10 client, and a SQL server. Then, there is the attacker machine which resides outside of the domain. Each of these systems will be explained in detail, as well as the role that they play within the Kerberos Authentication protocol and ticket forgery attacks.

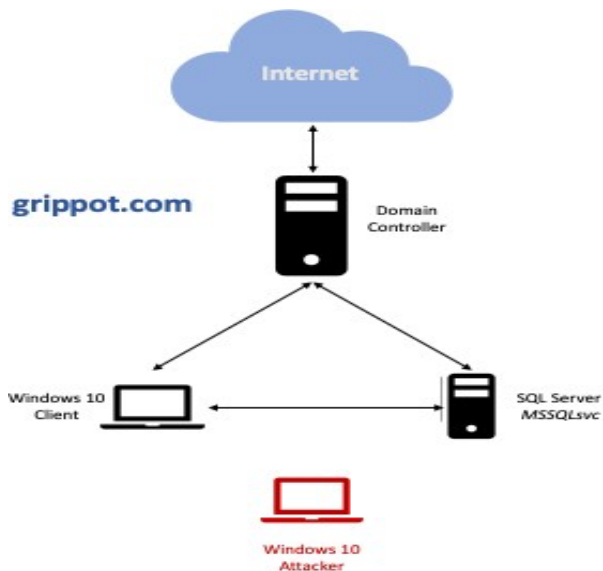

*Figure 10: The Virtual Environment*

This virtual environment was created using VirtualBox [8] as the hypervisor. VirtualBox is a type-2 hypervisor, meaning that it runs as software on the host OS. It can be downloaded from the website at https://www.virtualbox.org/. Each of the systems in this environment are running on separate VirtualBox VMs.

The domain controller (DC) is the most important system in the environment and serves many purposes. It is a Windows Server 2019 machine with two network interface cards (NIC), critical for the functioning of this environment. These can be seen in *Figure 11* below.

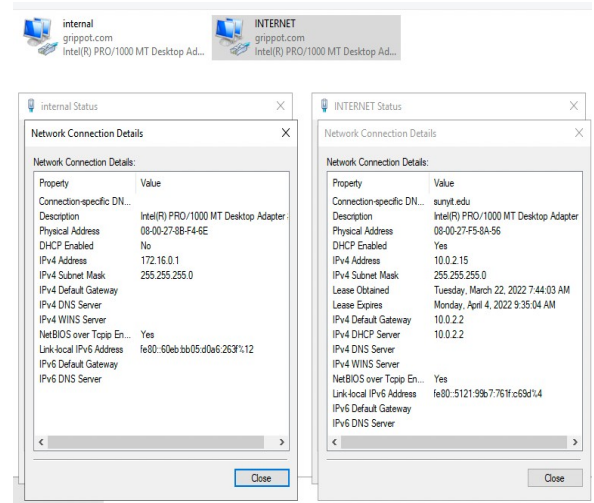

*Figure 11: Network Interfaces for DC*

One of the NICs is connected to the Internet from the host machine running VirtualBox. The other NIC is internal and has a static IP address of 172.16.0.1 as shown above. The reason for the domain controller to have two NICs is because in this environment it is also acting as a router. This means that not only is the domain controller running Windows Active Directory, but it is also routing traffic for all the other VMs. In order for this to be possible, the domain controller also must be running DNS and DHCP as shown in *Figure 12*, the Windows Server Manager.

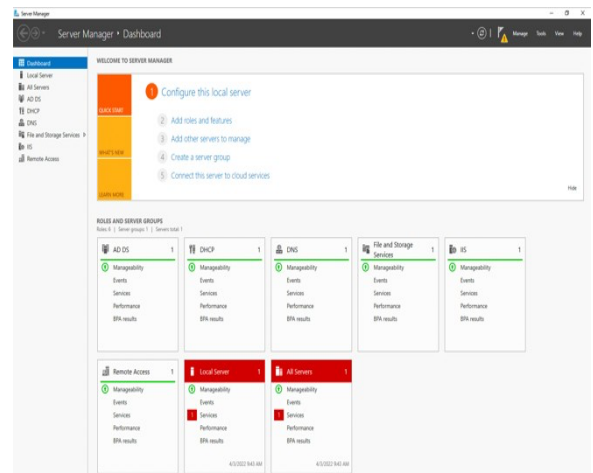

*Figure 12: Windows Server Manager for DC*

Having these services configured, the DC will be able to dynamically assign IPv4 addresses within the scope as shown in *Figure 13*.

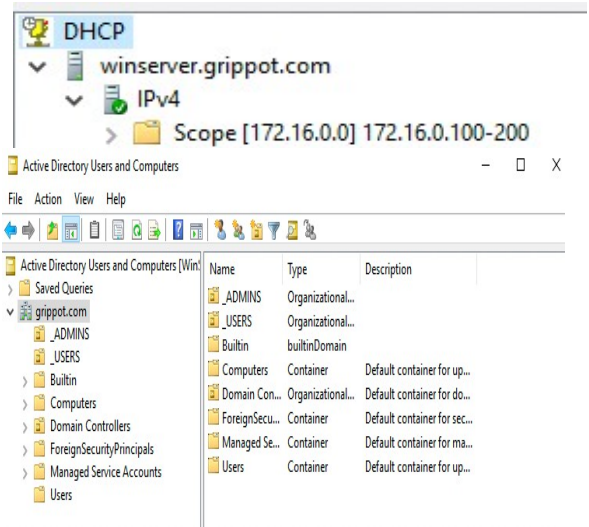

*Figure 13: DHCP on DC*

It will also act as the DNS server for its clients, and it will be able to route traffic to the Internet. Most importantly, however, the DC is running Windows Active Directory, and will act as the KDC for Kerberos. It is responsible for the *grippot.com* domain, which can be seen in *Figure 14*.

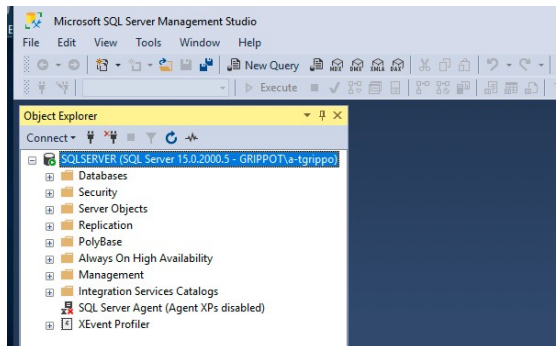

*Figure 14: The grippot.com Domain*

The *_ADMINS* and *_USERS* organizational units were created for the experiment. All the other containers were created by default when AD was configured on this server.

The next system in this environment is the Windows 10 client. This is simply a Windows 10 VM that is a part of the *grippot.com* domain.

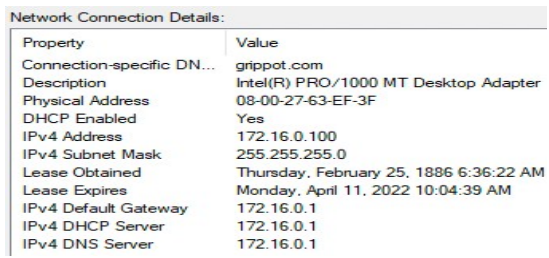

*Figure 15: Windows Client*

It has one internal NIC (*Figure 15*) that is connected to the domain controller. This is the machine that will act as a legitimate user that wants to access the SQL server within the domain.

The SQL server is another Windows Server 2019 machine with an internal NIC similar to the client.

*Figure 16: Microsoft SQL Server Management Studio*

It is running Microsoft SQL Server 2019 with Windows Authentication mode enabled to allow for Kerberos authentication for clients in the domain. Clients can connect to this SQL server either with the Microsoft SQL Server Management Studio as shown above in *Figure 16*, or they can use the *sqlcmd* utility to access it through a command-line interface. They will connect to the server through TCP 1433, which is the most common port for SQL. It is important to note that this service is not inherently vulnerable to ticket forgery attacks. This is because of the default service account that runs the MSSQLsvc service.

*Figure 17: MSSQLsvc*

The Log on service account for MSSQLsvc must be changed for the SQL server to be vulnerable to ticket forgery attacks. As shown in *Figure 17*, it has been changed to the SQLServiceAcc user in Active Directory, which has a weak password that we will be able to crack. This means that TGS tickets for the SQL server will be encrypted with the password of the SQLServiceAcc's password which is set to *Password123*.

Lastly, the attacker machine is running Windows 10. It is part of the local area network and is connected to the DC through its internal NIC. This could emulate an attacker who has gained access to the network, or an attacker who has compromised a Windows 10 machine within the domain. This is the machine that will be used to forge Kerberos tickets.

### IV. EXPERIMENT

An important point to understand about ticket forgery attacks in Active Directory is that they are a post-exploitation technique. This means that an attacker already has compromised a machine within the network and is looking to escalate their privileges and pivot to other machines. For example, an attacker may exploit some vulnerability in a machine and create a reverse shell or command and control (C2) channel so that they can control it from their own machine. They would then look for other machines on the network that they can potentially access and attempt to forge Kerberos tickets to gain control over more resources.

The most popular tool for forging Kerberos tickets, and for Kerberos vulnerabilities in general, is Mimikatz [9]. This is a popular open-source tool, that is used by both attackers and penetration testers to escalate privileges within a Windows network. It is a post-exploitation tool that would typically have to be downloaded onto the compromised machine.

Alternatively, attackers who have established a meterpreter shell on the compromised host can use the Mimikatz script through Metasploit without having to download it on the disk of the host [10]. Given how popular Metasploit is, this is the most common workflow.

This experiment is focused on the forged Kerberos tickets. Thus, we will not be going over any exploitation techniques with Metasploit or similar frameworks. Since we created this environment, we have the advantage of knowing all the credentials for the machines and controlling their security mechanisms. Leveraging this, we will skip over the initial exploitation phase, and get right into Mimikatz and forging Kerberos tickets using the Windows 10 attacker machine.

*A. Silver Ticket Attack*

To review, the silver ticket attack involves forging a TGS ticket for a specific service in the network. In this case, the MSSQLsvc service on the SQL server machine. For this to be successful, an attacker would need the password or NTLM service hash of the service account to properly encrypt the TGS ticket. To perform the attack, we will need to crack this hash to get the password so that we can use it to forge tickets.

*1) Download the Mimikatz Tool:* We will first download the Mimikatz tool from https://github.com/gentilkiwi/mimikatz/releases onto the attacker machine and extract the contents of the ZIP file. The executable file \x64\mimikatz.exe will be used to forge the TGS ticket. It is important to note that Windows Defender Antivirus will flag this download as malicious and must be disabled before downloading.

*2) Obtain the Service Account Password:* To get the password of the service account that will be used to encrypt the TGS, an attacker would first need to steal a legitimate TGS ticket. There are several ways to do this such as sniffing the network traffic in the local area network and hoping that a legitimate user requests the TGS [11]. For the simplicity of this lab, we will just take the real TGS ticket from the Windows 10 client machine. Kerberos tickets are stored in a cache in memory and can be viewed with the *klist* command. To extract these tickets from memory, we can use the Mimikatz command *kerberos::list /export* as shown in *Figure 18*.

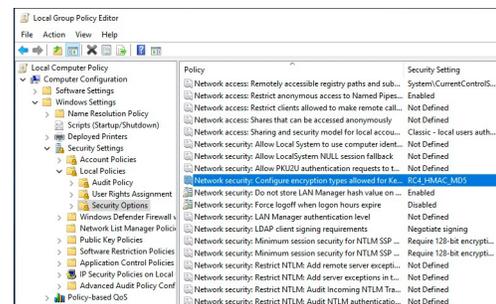

*Figure 18: Exporting Kerberos Tickets with Mimikatz*

The TGS ticket on the bottom is the one that we want. As seen above, it is for sqlserver.grippot.com in the GRIPPOT.COM domain and used by the *bross* user account which is currently logged in to the Windows 10 client. It is encrypted with RC4 and is valid for 10 hours. Mimikatz will store each of these tickets as files in the current working directory. We can then copy the TGS ticket over to the attacker machine VM.

Once the attacker has stolen the TGS ticket, they can then begin to crack the password used to encrypt the ticket through a technique called kerberoasting. Kerberoasting is a technique where an attacker attempts to brute force the password of a service account offline [12]. This is possible because all the tickets in the domain are encrypted with the RC4 algorithm which is weaker than the default AES256.

*Figure 19: Local Group Policy Settings on Domain Controller*

The supported Kerberos encryption algorithms can be changed in the Local Group Policy Settings as shown in *Figure 19*. This setting must be changed on all computers in the domain.

To crack the hash, we will use the Kerberos tool kit which can be found on GitHub [13]. Specifically,

the *tgsrepcrack.py* program will be used. The command to do this is shown below in *Figure 20*.

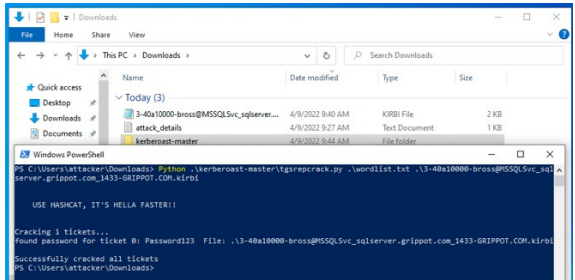
*Figure 20: Kerberoasting the Hash*

We give the program a custom wordlist that contains several common passwords to use. One of which is the correct password, *Password123*. We also supply the program with the legitimate TGS ticket that was stolen from the Windows 10 client machine. As seen in the screenshot, the password was successfully cracked and is displayed on the screen.

*3) Forge the TGS Ticket:* Now that we have the password for the SQLServiceAcc user, which is running the MSSQLsvc service, we can begin to forge the silver ticket. Mimikatz will be used to do this, which has been downloaded onto the attacker machine. To forge a silver ticket, Mimikatz requires several arguments. First, you need the SID or security identifier of the domain. This is a string that is unique to the domain [14]. Each user account in Active Directory is identified by their SID with a RID or relative identifier concatenated at the end. Mimikatz will also need the name of the domain, the fully qualified domain name (FQDN) of the target, the name of the target's service, the RC4 hash of the password, and the name of the user to impersonate.

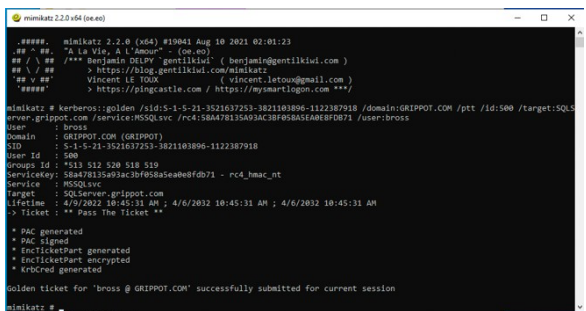
*Figure 21: Creating the Silver Ticket*

The full command is shown above in *Figure 21*. Mimikatz has successfully created the forged TGS ticket for the *bross* user. As specified by the /ptt option, Mimikatz has automatically stored the ticket in the Kerberos cache in memory.

*4) Access the SQL Server:* With the TGS ticket in the cache, the attacker machine can now access the MSSQLsvc as the *bross* user.

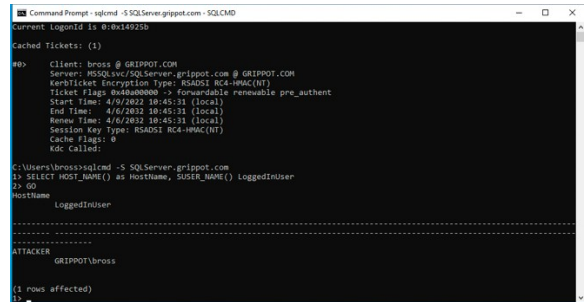
*Figure 22: Successful Authentication to MSSQLsvc*

Using the *sqlcmd -S SQLServer.grippot.com* command, the attacker can successfully access the MSSQLsvc. In *Figure 22*, the SQL query shows that the service believes the logged-in user is *bross*. The forged TGS ticket is also shown above using the *klist* command.

*B. Golden Ticket Attack*

Instead of the TGS ticket, the golden ticket attack aims to forge the TGT ticket. With a valid TGT ticket for a user with elevated privileges, an attacker can have great control over the domain. While this attack is more dangerous than the silver ticket attack, it is harder to perform in the real world. This is because an attacker would need to obtain the krbtgt account's NTLM hash. This would involve compromising an account with elevated privileges in order to dump hashes of accounts within the domain.

To see this attack in action, we will use the domain controller at *WinServer.grippot.com*, the Windows 10 client, and the Windows 10 attacker machine which is not a part of the grippot.com domain.

*1) Obtain the krbtgt NTLM Hash:* We will be using the Mimikatz tool to obtain the krbtgt NTLM hash. Mimikatz has a *dcsync* command that takes advantage of the Directory Replication Service Remote Protocol, or MS-DRSR [15]. This is a service used for communication between domains so that they can replicate information between each other [15]. By using the *dcsync* command in Mimikatz, we are essentially pretending to be a domain controller, and asking for the password data of the krbtgt account. However, for this to be possible, we need the proper permissions.

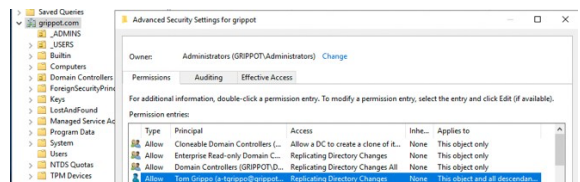
*Figure 23: Replicating Directory Changes Permission*

In *Figure 23* we can see that the *a-tgrippo* account has the Replicating Directory Changes permission. This will allow us to use the *dcsync* command.

From the Windows 10 client machine, we can use Mimikatz as the *a-tgrippo* account to dump the hash of the krbtgt account. This would simulate a compromised Windows machine within the domain that an attacker may utilize.

Figure 24: Obtaining krbtgt Hash

The command shown above is used to dump the NTLM hash along with other information pertaining to the *krbtgt account.* We will then copy this hash to the Attacker machine which is not part of the *grippot.com* domain. From there we can begin to forge the golden ticket.

*2) Forge the TGT Ticket:* Now that the attacker has the krbtgt account hash, they can forge the TGT. We will use Mimikatz again to do this. It will require the targeted domain, the security identifier of the domain, the krbtgt hash, and the user that you wish to impersonate. In this case, we will be impersonating the Administrator account with the relative identifier of 500.

Figure 25: Forging a TGT with Mimikatz

The Mimikatz command is shown in *Figure 25* above. The /ptt option is also passed so that the TGT is automatically copied to our cache in memory. With this ticket, we will be able to request TGS tickets from the domain controller so that we can access resources within the domain. Since we are impersonating the Administrator account, we can access the C: drive of the domain controller. This is shown in *Figure 26* below.

Figure 26: Using the TGT

Using the TGT, we can access files within the domain controller from a machine that is not a part of the domain. More importantly, at the top of the screenshot in *Figure 26* we can see that the TGT is valid for 10 years. This is the default expiration time for Mimikatz, and it will allow the attacker greater persistence in the domain.

*C. Detection*

Detecting these types of attacks can be difficult, especially the silver ticket attack. The easiest way to gain visibility into the Kerberos authentication process is by looking at the Windows Security Event Logs. There are several logs that pertain to Kerberos including 4624, 4634, 4672, and 4769. The first three event IDs are related to login events. For instance, 4624 is the log generated when a user successfully logs in to a system [16]. By closely analyzing these security event logs it is possible to catch a forged ticket in action.

For the golden ticket attack in this experiment, we can see a 4769 security event generated on the domain controller.

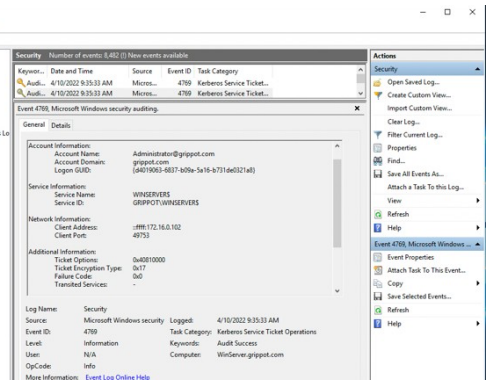

Figure 27: Windows Security Event 4769

This is for the request of a ticket granting service ticket. As shown in *Figure 27*, the ticket is requested by the Administrator account on 4/10/22 at 9:35 am. There are a couple of things that are suspicious about this log. First, there is no hostname for the client. If the client were a part of the domain, their hostname would appear in the log. Instead, just the IP address is shown. Also, more importantly, there is no record of a 4768 security event. This is the event that logs the request of a TGT. This could be a good indicator that the user has forged their own TGT, and never retrieved one from the domain controller.

Since the silver ticket attack does not involve the domain controller, it is harder to detect. In fact, for this experiment, it was not detected at all. The only significant logs generated were on the SQL server for login (4624) and log out (4634). Since we impersonated the *bross* user account for the silver ticket attack, these logs look normal. Typically, though, if an attacker is using a username that doesn't exist, these logs can be useful.

## V. ANALYSIS AND DISCUSSION

Although we weren't as successfully detecting these attacks in the virtual environment, there are several ways in which ticket forgery can be detected in the real world.

Many of these detection methods involve analyzing the tickets to look for anomalies. This can include things such as non-existent usernames, weaker encryption types, and suspicious privileges. The problem with this is that a smart attacker can easily correct these parameters and forge tickets that look completely legitimate. The only parameter that an attacker would be unlikely to change is the ticket lifetime. The default MaxTicketAge for legitimate tickets is 10 hours, whereas tickets created with Mimikatz have a default of 10 days [17]. We can see this in *Figure 25* with the golden ticket that we created. Its lifetime is from 4/10/2022 to 4/7/2032. This also holds true for the silver ticket created in *Figure* 21. An attacker could easily change the lifetime to 10 hours to be stealthier, however, they likely wouldn't. This is because forged Kerberos tickets, and forged TGTs in particular, are used for persistence within a network. Therefore, an attacker would want their ticket to last as long as possible so that they can continue to gain access to systems even after they are kicked off. Thus, a detection method that looks at the MaxTicketAge parameter of Kerberos tickets would be effective in detecting these types of attacks.

Overall, it would be advisable to implement a centralized log management system [18] so that all the logs generated across the network are stored in one place. Since ticket forgery attacks often generate logs across many systems, this would make analysis easier.

## VI. CONCLUSION

As long as organizations continue to employ Active Directory Domain Services, ticket forgery attacks against the Kerberos authentication protocol will remain a threat. These attacks are complex and require a deep understanding of both Active Directory and Kerberos. They exploit vulnerabilities within the Kerberos authentication protocol and allow potential attackers to impersonate users and privileges within the network. The golden ticket attack involves forging a TGT that can be used to access any services on the network. The silver ticket attack involves forging a TGS ticket that can be used to access a specific service on the network. To demonstrate these attacks, we created a virtual environment that simulated a real organization. This included a domain controller to run Active Directory Domain Services, a SQL server, and client machines. We then performed these attacks within the virtual environment to observe their functioning.

Specifically, the Windows security logs were analyzed to detect the forged tickets in the network. As shown, these attacks are difficult to detect once the tickets have been forged. This is especially true for the silver ticket attack which does not need to connect to the domain controller.

Regardless, both attacks are dangerous and should be recognized as a serious threat by corporations. It would be advisable to implement a powerful log collecting service to gain additional visibility into the network and catch forged tickets before they cause serious damage.

Future work includes integrating the proposed approaches with our cybersecurity framework [23-57] to detect the silver ticket attack in the 5G networks.